\documentstyle[12pt]{article}
\newcommand{\beq}{\begin{equation}}
\newcommand{\eeq}{\end{equation}}

\def\op{operator}
\def\dop{Dirac operator}
\def\tly{topologically}

\def\zm{zero mode}

\begin{document}
\begin{titlepage}
\begin{flushright}
NBI-HE-93-53 \\
September 1993\\
\end{flushright}
\vspace{0.5cm}
\begin{center}
{\large {\bf An $O(3)$ Global Anomaly in 0+1 Dimension}}\\
\vspace{1.5cm}
{\bf Minos Axenides}
\footnote{e-mail:axenides@nbivax.nbi.dk}\\
\vspace{0.4cm}
{\em The Niels Bohr Institute\\
University of Copenhagen, 17 Blegdamsvej, 2100 Copenhagen, Denmark}\\
\vspace{0.4cm}
{\bf Andrei Johansen}
\footnote{e-mail:johansen@lnpi.spb.su}\\
\vspace{0.4cm}
{\em The St.Petersburg Nuclear Physics Institute\\
 Gatchina, St.Petersburg District, 188350 Russia}\\
\vspace{0.4cm}
{\bf Holger Bech Nielsen}
\footnote{e-mail:hbech@nbivax.nbi.dk}\\
\vspace{0.4cm}
{\em The Niels Bohr Institute\\
University of Copenhagen, 17 Blegdamsvej, 2100 Copenhagen, Denmark}\\

\end{center}
\begin{abstract}
We present a simple exactly solvable
quantum mechanical example of the global anomaly
in an $O(3)$ model with an odd number
of fermionic triplets coupled to a gauge field
on a circle.
Because the fundamental group is non-trivial,
$\pi_1 (O(3)) ={\bf Z}_2$,
fermionic level
crossing - circling occurs in the eigenvalue spectrum of
the 1-dim. Dirac \op under continuous external field transformations.
They are shown to be related to the presence of an odd number
of normalizable zero modes in the spectrum of an appropriate
2-dim. Dirac \op . We argue that fermionic degrees of freedom
in the presence of an infinitely large external field violate
perturbative decoupling.

\end{abstract}
\end{titlepage}
\newpage

The phenomenon of fermion level crossing in field theory
and its relation to
the existence of normalizable fermionic zero modes of the \dop \ in
the presence of \tly \ non-trivial gauge fields is well known \cite{sc}.

The $SU(2)$ global anomaly \cite{witten}
in particular is related to the phenomenon of
level crossing in the eigenvalue
spectrum of the 4-dim. ($D=4$) Dirac operator.
This eigenvalue flow
corresponds to the existence of a \zm \ for an appropriately defined
5-dim. ($D=5$) \dop \
in the presence of an external topologically
non-trivial gauge field. As a result the fermionic path integral
is not gauge invariant and the theory is not self-consistent.
Recently a generalization was presented \cite{mah}
for the case of an $SU(2)$ theory with an odd number of Weyl doublets
and arbitrary Yukawa couplings and quark masses.

As a result of the antisymmetric character of the $D=4$ Dirac \op \
its spectrum of eigenvalues is organized in paires $\lambda, -\lambda$
on the complex plane.
Under topologically non-trivial gauge transformation a fermionic level
circling effect was shown to occur for an odd number of positive
eigenvalues.
This in turn was related to the existence of an odd number of
normalizable \zm s in the spectrum of the $D=5$ \dop \
consistently with the Atiyah-Singer index theorem.
While the general picture appears to be unambiguous some
subtle question still remain with regard to the precise nature of
fermionic eigenvalue rearrangment on the complex plane,
consistently with the index theorem.

Of equal interest is also the issue of pertubative decoupling
of heavy fermions from the light sector of a theory in the presence of
Yukawa interactions and global
anomalies \cite{banks}. More specifically it was argued, in the context
of effective field theories with $U(1)$ global anomalies that the low
energy physics is in fact sensitive to the presence of otherwise
integrated out heavy fermions through their zero modes(nonperurbative
non-decoupling). We would like to investigate this possibility for the
case of the $SU(2)$ global anomaly as well.

In this paper we examine a simple quantum mechanical model
which captures, we believe, the essential properties of the $D =4$
$SU(2)$ gauge field theory with odd number of fermionic Weyl doublets
that posseses an anomaly\cite{witten,mah}.
This is an $O(3)$ gauge model with odd number of
real fermion $\psi$ in the
fundamental representation of the gauge group.
With no loss of generality we first consider the case of a
single real fermionic
triplet.
A generalization to the case of an odd number of fermionic triplets
is of course straightforward.
The lagrangian reads as
\beq
L =i \psi \dot{\psi} +\psi A \psi,
\eeq
where $A$ is an $O(3)$ gauge field and $\dot{\psi}$ means the
derivative in time variable $t.$
We shall consider the compactified version of the theory where
the 'fields' live on a circle, so that $0 \leq t \leq \beta$
where $\beta$ is a positive parameter.
We also introduce appropriate boundary conditions for fields
\beq
\psi (\beta) =- \psi (0), \;\;\;
A(\beta) = A(0).
\eeq
This model has a gauge invariance under transformations
\beq
\psi \to e^{i\alpha} \psi ,\;\;\;
A \to  e^{i\alpha} i\partial_t e^{-i\alpha} .
\eeq
where $\alpha$ is a periodic function taking values in the $O(3)$
algebra.
It is clear that the gauge field $A$ has almost no physical degrees
of freedom.
Indeed by an appropriate gauge transformation it may take the form of a
constant matrix belonging to the $O(3)$ algebra.
Subsequently by a constant $O(3)$ rotation it can be transformed to
\beq
A = a T_2 ,
\eeq
where $a$ is a real number and $T_2$ is a generator of $O(3)$ group.
If $a$ is not a multiple of $2\pi /\beta$ then this gauge field
can not be removed by a gauge transformation with a parameter
periodic in time.
In turn if $a =  2\pi n /\beta$, $n \in {\bf Z}$, then this
gauge field is gauge equivalent to $a =0 .$
It is also clear that the spectrum of the 'Dirac' \op \
\beq
iD =i\partial_t +A
\eeq
does not change under a shift $a\to a+ 2\pi /\beta .$

However because of the homotopic theorem \cite{hu}
\beq
\pi_1 (O(3)) ={\bf Z}_2
\eeq
and in analogy with Witten's $SU(2)$ anomaly
we expect our quantum mechanical
model be equally inconsistent.
This is indeed the case, as we will see below.

First let us note that the 'Dirac' operator $D$
is antisymmetric and antihermitian
\beq
D^T = -D, \;\;\; D^+ = -D.
\eeq
The eigenvalues of this \op \ are purely imaginary.
Moreover the non-zero eigenvalues $i\lambda$
are paired as $(i\lambda , -i\lambda) .$
We shall soon see that for a generic gauge field this \op \ has no
zero modes.

The partition function of our model is proportional
to the Pfaffian of the \op \ $D .$
Let us define this Pfaffian as a square root of the
determinant of the 'Dirac' \op \ $D$
\beq
{\rm Pf} D = (\det D)^{1/2}.
\eeq
For a generic gauge field this square root can be defined as
a product of all positive eigenvalues.

Now let us find the spectrum of the 'Dirac' operator $D$ in an
external gauge field $A = a T_2 $
for an abitrary real constant $a .$
The equation for the eigenfunction that corresponds to an eigenvalue
$i\lambda$ reads as
\beq
D\psi =i\lambda \psi.
\eeq
The solution is given by
\beq
\psi = e^{it\lambda} e^{-it aT_2} \psi_0 ,
\eeq
where $\psi_0$ is a constant 3-component vector.
The spectrum is determined by imposing the antisymmetric
boundary condition to the fermionic field
\beq
e^{i\beta\lambda} e^{-i\beta aT_2} \psi_0 = - \psi_0 .
\eeq
The existence of a non-zero solution $\psi_0$ to the latter equation
is equivalent to
\beq
\det \left( e^{i\beta\lambda} e^{-i\beta aT_2} \psi_0 +1 \right) =0 .
\eeq
{}From this equation we find three types of eigenvalues (Fig.1)
($+, 0, -$) given by
\beq
\lambda^{(0)}_k = (\pi +2\pi k)/\beta ,\;\;\;
\lambda^{(+)}_k = a+ (\pi +2\pi k)/\beta ,\;\;\;
\eeq
$$\lambda^{(-)}_k = -a + (\pi +2\pi k)/\beta ,
\;\;\;\; k \in {\bf Z} .$$

The spectrum of the Dirac operator is organized in pairs
$(\lambda,-\lambda)$ of eigenvalues with opposite sign and type. As it
is also depicted in fig.2 when we adiabatically change $a$
into $a+2\pi/\beta$ there exists exactly two eigenvalues
$(\lambda^{+}_{3},-\lambda^{-}_{3})$ that cross the zero of the real
eigenvalue axis and into the $(\lambda^{+}_{1},-\lambda^{-}_{1})$ levels
respectively preserving their type. All the other eigenvalue pairs of
the positive and negative eigenvalue type rearrange themselves
accordingly into distinctly different available levels of the same type
and sign character. The eigenvalues of the zero type dont feel the
external field and thus they intact.

It is easy to calculate the value of the Pfaffian. We find it to be
\beq
{\rm Pf} D = \prod_{k=0}^{\infty} \left( a + \frac{\pi + 2\pi k}{\beta}
\right)
\left( -a + \frac{\pi + 2\pi k}{\beta}
\right)
\left(\frac{\pi + 2\pi k}{\beta}
\right)=
\eeq
$$= {\rm const}
\prod_{k=0}^{\infty} \left( 1-  \left( \frac{a\beta}{\pi + 2\pi k}
\right)^2
\right) \sim \cos a\beta/2 .$$
It is thus obvious that when $a$ changes adiabatically to $a +
2\pi/\beta$ the Pfaffian changes sign.

At this point we remark that in order for our model to make sense
as a quantum gauge theory we have to integrate over all gauge
field configurations.
To that end we choose a gauge for $A =a T_2 .$
Because the global anomaly breaks the gauge invariance under
``big'' gauge rotation $a \to a+2\pi/\beta$
we must integrate over the whole real line.
We may observe that the partition function
$$Z=\int^{\infty}_{-\infty} da \; \cos \frac{a\beta}{2} =0$$
vanishes identically, a manifestation of the global anomaly.

It is interesting now to look at the case when $a$ is complex.
In this case
the 'Dirac' \op \ becomes non-hermitian and hence its eigenvalues
are complex (Fig. 1 $\&$ 3).
Eqs. (13) suggest that as the parameter $a$ changes adiabatically
to $a = 2\pi/\beta $ the  pair of eigenvalues
$(\lambda^{+}_{3},\lambda^{-}_{3})$ that previously crossed the zero real
axis in this case bypass each other at a distance (``level circling'')
which is proportional to twice the imaginary part of the external field
$a$.

We can find a corresponding normalizable \zm \ of an appropriate
euclidean $D=2$ Dirac \op .
The natural candidate reads as
\beq
\sigma_1\partial_t +  \sigma_2 \partial_x -i\sigma_1 A ,
\eeq
where $x$ is a second coordinate parametrizing a continuous
change of the gauge field (the $D=2$ field is taken in the
'Hamiltonian' gauge $A_x =0$, $A_t =A$).
Multiplying this equation by $i\sigma_1$
we get the following equvalent form
\beq
\hat{D}=i\partial_t -  \sigma_3 \partial_x + A .
\eeq
This equation splits into two equations for
the components of a spinor $\psi =(\psi_1 ,\psi_2) .$
However it is clear that we should take only one of these equations
to describe the process of level crossing.
Thus we choose
\beq
(\partial_z - A) \psi_1 =0,
\eeq
where $z=x+it .$
It is clear that the \op \ $\partial_z - A$ is antisymmetric
and hence its non-zero eigenvalues are paired.
Hence the number of zero modes of the \op \ $\partial_z - A$
is invariant under deformations of the gauge field.
Let us take
\beq
a(x) =a+ \frac{2\pi}{\beta} ({\rm th} x + 1).
\eeq
Then the field $A$ interpolates between values $a$ at $x\to -\infty$
and $a + 2\pi/\beta$ at $x\to +\infty$
and hence it plays the role of an instanton.
It is easy to check that there is a unique solution
to the above equation which is normalizable and periodic
in the ``time'' variable $t.$
For $0\leq a < \pi/\beta$ we have
\beq
\psi = ({\rm cosh}x)^{-\frac{\pi}{\beta}} e^{-ax
-it\pi/\beta} \psi_{0},
\eeq
where $\psi_{0}$ is the $O(3)$ triplet given by (up to a normalization
factor)
\beq
\psi_{0} = (1 , -i , 0) .
\eeq
In the case $\pi/\beta < a\leq 2\pi/\beta$ we get
\beq
\psi = ({\rm cosh}x)^{-\frac{\pi}{\beta}} e^{(2\pi/\beta-a)x
-it3\pi/\beta} \psi_{0}.
\eeq
Finally when $a=\pi/\beta$ there is no
anti-periodic in time $t$ normalizable \zm .
This case corresponds to the absence of level crossing.
Indeed there is a double degeneracy of levels $\lambda^{\pm} = 2\pi k/\beta$,
$k\in {\bf Z}$, in the spectrum of the $D=1$ \dop .
As a result its determinant vanishes at this value of the
gauge field due to the contribution $k=0 .$
Hence its sign does not change under a ``big''
gauge rotation.
In turn in an analogy with Witten's anomaly the existence of the
normalizable \zm \ for $a\neq \pi /\beta$
is related to the level crossing (circling)
of the $D=1$ 'Dirac' \op .

This completes our demonstartion for the existence of a \zm \
for the $D=2$ \dop .
As stated before in an $O(3)$ model with an odd number of fermionic
triplets we would expect to find an odd number of \zm s for
the $D=2$ \dop .
This would be related to a level crossing/circling of an odd number of
eigenvalues in the spectrum of the $D=1$ \dop .

We proceed now to examine a healthy version of our model
with an even number, say two, of fermionic triplets.
We introduce a ``scalar''
field $\Phi$ which transforms under gauge rotations parametraized by
$\alpha$ (an element of the $O(3)$ algebra)
\beq
\Phi \to e^{i\alpha} \Phi e^{-i\alpha}.
\eeq
Our lagrangian is given by
\beq
{\cal L} =
i\psi \dot{\psi} + \psi A \psi +
i\chi \dot{\chi} + \chi (A+\Phi) \chi .
\eeq
In this theory
because the gauge invariance is intact we must integrate in
the partition function over the gauge field.
It is trivially seen that the model is indeed healthy as
$$Z= \int^{2\pi/\beta}_{0} da \; \cos^2 \frac{a\beta}{2} = \pi/\beta .$$

Let us now explore the limit where $\Phi \to \infty$
and a perturbative decoupling of $\chi$ field is expected.
Due to gauge invariance we take $A+\Phi \sim h T_2$
($T_2$ is an $O(3)$ generator and $h$ is a constant).
If we let $\Phi \to \infty$ then $h \to \infty$ two components of
the $\chi$ fermionic triplet become ``heavy''.
{}From eq.(13) we observe that the third $\chi$ component
remains light and free.

By integrating out the $\chi$ field in the functional integral
we get a factor $\sim \cos h\beta/2 .$
When we let $\Phi \to \infty$ ($h\to \infty$) this limit
does not exist.
At this point we may draw two independent conclusions:

Firstly our initially healthy model of two fermionic triplets
in the ``decoupling'' limit where one of them ($\chi$) becomes
infinitely massive remains healthy.
This is because level crossing/circling in the spectrum of the
$D=1$ \dop \ for the perturbatively decoupled $\chi$
still occurs.
In other words we observe a non-perturbative non-decoupling of
the heavy $\chi$ field.

Secondly, the non-existence of the limit $h\to \infty$
also implies that there is no well defined low energy effective
lagrangian for the $\psi$ field alone.
In the realistic case of the standard electro-weak theory
with an infinitely heavy top-quark doublet we may infer that
there is no well defined low-energy theory.

One of us A.J. acknowledges the high energy group at NBI for its
hospitality. The present work was supported in part by a NATO grant
GRG 930395. M.A. is grateful to the Carlsberg Foundation for financial
support.

\newpage
\subsection*{Figure Captions}
\vskip 6pt

$\;\;\;\;\;$Fig.1.
The spectra of the Dirac \op \ in the external gauge field
$A =a T_2$ in two cases:
1) $a=$ real and 2) $a=$ complex.
The square root of the determinant is defined respectively as a product
of eigenvalues which 1) are positive and 2) have positive real part.

Fig.2. The flow of real eigenvalues ($\equiv$ level crossing) as
the gauge field varies from $A = aT_2$ to
$A^U = (a + 2\pi/\beta) T_2$ ($2\pi/\beta \equiv$ radius of the circle).
Solid lines indicate the eigenvalues that define the determinant.

Fig.3. The flow of complex eigenvalues is depicted on the complex
$\lambda$-plane. The complex gauge field $A$
varies from $aT_2$ to $(a + 2\pi/\beta) T_2$ .
Solid lines indicate eigenvalues that define the determinant.
Level circling occurs between levels $\lambda_3$ and $-\lambda_3 .$
\end{document}